\documentclass[aps,twocolumn,prd,nofootinbib,superscriptaddress,showpacs]{revtex4}
\usepackage{txfonts}
\usepackage{graphicx}
\usepackage{dcolumn}
\usepackage{bm}
\usepackage{amssymb}
\usepackage{latexsym}

\newcommand{\be}{\begin{equation}}
\newcommand{\ee}{\end{equation}}
\newcommand{\bq}{\begin{eqnarray}}
\newcommand{\eq}{\end{eqnarray}}

\begin{document}


\title{Reconstructing quintom from WMAP 5-year observations: Generalized ghost
condensate}

\author{Jingfei Zhang}
\affiliation{Department of Physics, College of Sciences,
Northeastern University, Shenyang 110004, China}

\author{Yuan-Xing Gui}
\affiliation{School of Physics and Optoelectronic Technology, Dalian
University of Technology, Dalian 116024, China}

\begin{abstract}
In the 5-year WMAP data analysis, a new parametrization form for
dark energy equation-of-state was used, and it has been shown that
the equation-of-state, $w(z)$, crosses the cosmological-constant
boundary $w=-1$. Based on this observation, in this paper, we
investigate the reconstruction of quintom dark energy model. As a
single-real-scalar-field model of dark energy, the generalized ghost
condensate model provides us with a successful mechanism for
realizing the quintom-like behavior. Therefore, we reconstruct this
scalar-field quintom dark energy model from the WMAP 5-year
observational results. As a comparison, we also discuss the quintom
reconstruction based on other specific dark energy ansatzs, such as
the CPL parametrization and the holographic dark energy scenarios.

\end{abstract}

\maketitle

\section{Introduction}\label{sec:intr}

Observations of high-redshift supernovae indicate that the universe
is accelerating at the present stage~\cite{SN} and this accelerating
expansion also has been confirmed by many other cosmological
experiments, such as observations of large scale structure (LSS)
\cite{LSS}, and measurements of the cosmic microwave background
(CMB) anisotropy \cite{CMB}. We refer to the cause for this cosmic
acceleration as ``dark energy,'' which is a mysterious exotic matter
with large enough negative pressure and whose energy density has
been a dominative power of the universe. The combined analysis of
cosmological observations suggests that the universe is consists of
about $70\%$ dark energy, $30\%$ dust matter (cold dark matter plus
baryons), and negligible radiation. The astrophysical feature of
dark energy is that it remains unclustered at all scales where
gravitational clustering of baryons and nonbaryonic cold dark matter
can be seen. Its gravity effect is shown as a repulsive force so as
to make the expansion of the universe accelerate when its energy
density becomes dominative power of the universe.

Although the nature and origin of dark energy are unknown, we still
can propose some candidates to describe the properties of dark
energy. The most obvious theoretical candidate of dark energy is the
cosmological constant $\Lambda$ (vacuum energy) \cite{Einstein:1917}
with an equation of state $w=-1$. The cosmological constant is
rather popular in researches of cosmology and astrophysics due to
its theoretical simpleness and its great success in fitting with
observational data. However, as is well known, the two fundamental
problems, namely the ``fine-tuning'' problem and the ``cosmic
coincidence'' problem \cite{coincidence}, still puzzle us. Theorists
have made many efforts to try to resolve the cosmological constant
problem, but all of these efforts turn out to be unsuccessful
\cite{cc}.

Also, there are other alternatives to the cosmological constant. An
alternative proposal to explaining dark energy is the dynamical dark
energy scenario. The dynamical dark energy proposal is often
realized by some scalar field mechanism which suggests that the
energy form with negative pressure is provided by a scalar field
slowly rolling down its potential. So far, a lot of scalar-field
dark energy models have been studied. The models such as
quintessence \cite{quintessence}, $K$-essence \cite{kessence},
phantom \cite{phantom}, tachyon \cite{tachyon} and ghost condensate
\cite{ghost1,ghost2} are all famous examples of scalar-field dark
energy models. In these models, the quintessence with a canonical
kinetic term evolves its equation of state in the region of
$w\geqslant -1$ whereas the model of phantom with negative kinetic
term can always lead to $w\leqslant -1$; the $K$-essence can realize
both $w>-1$ and $w<-1$, but it has been shown that it is very
difficult for $K$-essence to achieve $w$ of crossing $-1$
\cite{Vikman:2004dc}.

However, the analysis of the current observational data shows that
the equation of state of dark energy $w$ is likely to cross the
cosmological-constant boundary (or phantom divide) $-1$, i.e. $w$ is
larger than $-1$ in the recent past and less than $-1$ today. The
dynamical evolving behavior of dark energy with $w$ getting across
$-1$ has brought forward great challenge to the model-building of
scalar-field in the cosmology. Just as mentioned above, the
scalar-field models, such as quintessence, $K$-essence, phantom,
cannot realize the transition of $w$ from $w>-1$ to $w<-1$ or vice
versa. Hence, the quintom model was proposed for describing the
dynamical evolving behavior of $w$ crossing $-1$ \cite{quintom} with
double fields of quintessence and phantom. The cosmological
evolution of such model has been investigated in detail
\cite{twofield,quintom2}. For the single real scalar field models,
the transition of crossing $-1$ for $w$ can occur for the Lagrangian
density $p(\phi, X)$, where $X$ is a kinematic term of a
scalar-field $\phi$, in which $\partial{p}/\partial{X}$ changes sign
from positive to negative, thus we require nonlinear terms in $X$ to
realize the $w=-1$ crossing
\cite{ghost2,Vikman:2004dc,Anisimov:2005ne}. When adding a high
derivative term to the kinetic term $X$ in the single scalar field
model, the energy-momentum tensor is proven to be equivalent to that
of a two-field quintom model \cite{Li:2005fm}. It is remarkable that
the generalized ghost condensate model of a single real scalar field
is a successful realization of the quintom-like dark energy
\cite{Tsujikawa:2005ju,Zhang:2006qu}. What's more, a generalized
ghost condensate model was investigated in
Refs.~\cite{Tsujikawa:2005ju,Zhang:2006qu} by means of the
cosmological reconstruction program. For another interesting
single-field quintom model see Ref.~\cite{Huang:2005gu}, where the
$w=-1$ crossing is implemented with the help of a fixed background
vector field. Besides, there are also many other interesting models,
such as holographic dark energy model \cite{holo} and braneworld
model \cite{Cai:2005ie}, being able to realize the quintom-like
behavior.

In any case, these dark energy models including the dynamical dark
energy models have to face the test of cosmological observations. A
typical approach for this is to predict the cosmological evolution
behavior of the models, by putting in the Lagrangian (in particular
the potential) by hand or theoretically, and to make a consistency
check of models by comparing it with observations. An alternative
approach is to reconstruct corresponding theoretical Lagrangian, by
using the observational data. The reconstruction of scalar-field
dark energy models has been widely studied. For a minimally coupled
scalar field with a potential $V(\phi)$, the reconstruction is
simple and straightforward \cite{simplescalar}. Saini et al.
\cite{Saini:1999ba} reconstructed the potential and the equation of
state of the quintessence field by parameterizing the Hubble
parameter $H(z)$ based on a versatile analytical form of the
luminosity distance $d_L(z)$. This method can be generalized to a
variety of models, such as scalar-tensor theories
\cite{scalartensor}, $f(R)$ gravity \cite{frgrav}, $K$-essence model
\cite{Li:2006bx,Gao:2007ep}, and also tachyon model \cite{holotach},
etc.. Tsujikawa has investigated the reconstruction of general
scalar-field dark energy models in detail \cite{Tsujikawa:2005ju}.

In this paper, we will investigate the quintom reconstruction from
the Wilkinson Microwave Anisotropy Probe (WMAP) 5-year observations.
We will focus on the generalized ghost condensate model and will
reconstruct this quintom scalar-field model using various dark
energy ansatzs including the parametric forms of dark energy and
holographic dark energy scenarios. In particular, we will put
emphasis on a new parametrization form proposed by WMAP team in
Ref.~\cite{Komatsu:2008hk}.

The paper is organized as follows: In section \ref{sec:para} we
address the dark energy parametrization proposed in
Ref.~\cite{Komatsu:2008hk} and describe the corresponding analysis
results of the WMAP5 observations. In section \ref{sec:ghost} we
perform a cosmological reconstruction for the generalized ghost
condensate model from various dark energy ansatzs and the fitting
results of the up-to-date observational data. Finally we give the
concluding remarks in section \ref{sec:concl}.

\section{A new dark energy parametrization in WMAP5}\label{sec:para}

The distinctive feature of the cosmological constant or vacuum
energy is that its equation of state is always exactly equal to
$-1$. Whereas, the dynamical dark energy exhibits a dynamic feature
that its equation-of-state as well as its energy density are
evolutionary with time. An efficient approach to probing the
dynamics of dark energy is to parameterize dark energy and then to
determine the parameters using various observational data. One can
explore the dynamical evolution behavior of dark energy efficiently
by making use of this way, although the results obtained are
dependent on the parametrizations of dark energy more or less.

Among the various parametric forms of dark energy, the minimum
complexity required to detect time variation in dark energy is to
add a second parameter to measure a change in the equation-of-state
parameter with redshift. This is the so-called linear expansion
parametrization $w(z)=w_0+w'z$, where $w'\equiv dw/dz|_{z=0}$, which
was first used by Di Pietro $\&$ Claeskens \cite{DiPietro:2002cz}
and later by Riess et al. \cite{Riess:2004nr}. However, when some
``longer-armed'' observations, e.g. CMB and LSS data, are taken into
account, this form of $w(z)$ will be unsuitable due to the
divergence at high redshift. The most commonly used form of
equation-of-state, $w(z)=w_0+w_a z/(1+z)$, suggested by Chevallier
$\&$ Polarski \cite{Chevallier:2000qy} and Linder
\cite{Linder:2002et} (hence, hereafter, this form is called CPL
parametrization, for convenience), can avoid the divergence problem
effectively. It should be noted that this parametrization form has
been investigated enormously in exploring the dynamical property of
dark energy in light of observational data. However, this form
cannot be adopted as it is when one uses the CMB data to constrain
$w(a)$~\cite{Komatsu:2008hk}. Since this form is basically the
leading-order term of a Taylor series expansion, the value of $w(a)$
can become unreasonably too large or too small when extrapolated to
the decoupling epoch at $z_*\simeq 1090$ (or $a_*\simeq 9.17\times
10^{-4}$), and thus one cannot extract meaningful constraints on the
quantities such as $w_0$ and $w_a$ that are defined at the {\it
present epoch}.

In order to avoid this problem, a new parametrized form was proposed
by the WMAP team \cite{Komatsu:2008hk},
\begin{equation}
w(a) = \frac{a\tilde{w}(a)}{a+a_{\rm trans}} - \frac{a_{\rm
trans}}{a+a_{\rm trans}}, \label{eq:wz}
\end{equation}
(here, it is marked as ``WMAP5 parametrization'') where
\begin{equation}
\tilde{w}(a)=\tilde{w}_0+(1-a)\tilde{w}_a,
\end{equation}
and $a_{\rm trans}=1/(1+z_{\rm trans})$ is the ``transition epoch,''
and $z_{\rm trans}$ is the transition redshift. In this form, $w(a)$
approaches to $-1$ at early times and the dark energy density tends
to a constant value at $a<a_{\rm trans}$. The dark energy density
remains totally sub-dominant relative to the matter density at the
decoupling epoch. At late times, $a>a_{\rm trans}$, one recovers the
widely used CPL form~\cite{Linder:2002et}, $w(a)=w_0+(1-a)w_a$.

In ``WMAP5 parametrization'', the present-day value of $w$,
$w_0\equiv w(z=0)$, and the first derivative, $w'\equiv
\left.dw/dz\right|_{z=0}$, are chosen as the free parameters, in
stead of the $\tilde{w}_0$ and $\tilde{w}_a$.

\begin{figure}[htbp]
\begin{center}
\includegraphics[scale=0.75]{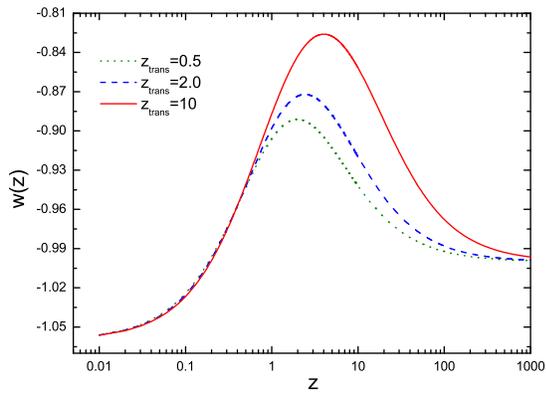}
\caption[]{\small The evolution of the equation of state of dark
energy, corresponding to the WMAP5 parametrization with $w_0=-1.06$
and $w'=0.36$, for the transition redshift $z_{\rm trans}=0.5$,
$2.0$ and $10$, respectively.}\label{fig:wz}
\end{center}
\end{figure}

In Ref.~\cite{Komatsu:2008hk}, the WMAP group constrains $w_0$ and
$w'$ in a flat universe from the WMAP distance priors ($l_A$, $R$,
$z_*$), combined with the Baryon Acoustic Oscillations (BAO) and the
Type Ia supernovae (SN) data. The results are that, for $z_{\rm
trans}=10$, the 95\% limit on $w_0$ is $-0.33<1+w_0<0.21$; the 68\%
intervals are $w_0=-1.06\pm 0.14$ and $w'=0.36\pm 0.62$. Note that
Ref.~\cite{Wang:2007mza} shows that the two-dimensional distribution
extends more towards south-east, i.e., $w>-1$ and $w'<0$, when the
spatial curvature is allowed. The evolutionary behavior of $w(z)$ is
plotted in Fig.~\ref{fig:wz}, using the best-fit results. It should
be noted that that Fig.~\ref{fig:wz} is slightly different from
Fig.~C1 of Ref.~\cite{Komatsu:2008hk} in that the revised best-fit
values, $w_0=-1.06$ and $w'=0.36$, are used in plotting this figure.

In this section, we have briefly introduced the new parametrization
of dark energy equation-of-state proposed by the WMAP team in the
5-year observations. The advantage of this parameterized form is
that the value of $w(a)$ is still reasonable when extrapolated to
the early times such as the decoupling epoch. We shall use this
parametrization with the observational constraint result to
reconstruct the generalized ghost condensate model in the next
section. As a comparison, we will also discuss other specific cases,
such as the CPL parametrization as well as the holographic dark
energy scenarios. This work is different from the previous ones
\cite{Tsujikawa:2005ju,Zhang:2006qu} in that the reconstruction is
implemented up to the decoupling epoch at $z_*\simeq 1090$.

\section{Generalized ghost condensate model and its
reconstruction}\label{sec:ghost}

As mentioned above, the dynamical dark energy can be realized by
some scalar-field mechanism. In particular, the quintom model was
proposed for describing the dynamical evolving behavior of $w$
crossing $-1$. The results of the current observational data
analysis show that the equation of sate of dark energy is likely to
cross $-1$, see, for example, Fig.~\ref{fig:wz}. So, it is necessary
to realize such a quintom behavior using some scalar field
mechanism. It is remarkable that the generalized ghost condensate
model is a successful single-real-scalar-field quintom model. In
this section, we shall focuss on the reconstruction of the
generalized ghost condensate model from the WMAP 5-year
observations. We will first briefly review the generalized ghost
condensate model of dark energy. Then, we will implement the
scalar-field dark energy reconstruction according to the WMAP5
parametrization. As a comparison, we will also perform the same
reconstruction program for other specific models such as the CPL
parametrization and the holographic dark energy scenarios.

\subsection{Generalized ghost condensate model}

First, let us consider the Lagrangian density of a general scalar
field $p(\phi, X)$, where
$X=-g^{\mu\nu}\partial_\mu\phi\partial_\nu\phi/2$ is the kinetic
energy term. Note that $p(\phi, X)$ is a general function of $\phi$
and $X$, and we have used a sign notation $(-, +, +, +)$.
Identifying the energy momentum tensor of the scalar field with that
of a perfect fluid, we can easily derive the energy density of dark
energy, $\rho_{\rm de}=2Xp_X-p$, where $p_X=\partial p/\partial X$.
Thus, in a spatially flat Friedmann-Robertson-Walker (FRW) universe
involving dust matter (baryon plus dark matter) and dark energy, the
dynamic equations for the scalar field are
\begin{equation}
3H^2=\rho_{\rm m}+2Xp_X-p,\label{hsqr}
\end{equation}
\begin{equation}
2\dot{H}=-\rho_{\rm m}-2Xp_X,\label{hdot}
\end{equation}
where $X=\dot{\phi}^2/2$ in the cosmological context, and note that
we have used the unit $M_P=1$ for convenience. Introducing a
dimensionless quantity
\begin{equation}
r\equiv E^2= H^2/H_0^2,
\end{equation}
we find from Eqs.~(\ref{hsqr}) and (\ref{hdot}) that
\begin{equation}
p=[(1+z)r'-3r]H_0^2,\label{p}
\end{equation}
\begin{equation}
\phi'^2p_X={r'-3\Omega_{\rm m0}(1+z)^2\over r(1+z)},\label{px}
\end{equation}
where prime denotes a derivative with respect to $z$. The equation
of state for dark energy is given by
\begin{equation}
w={p\over \dot{\phi}^2 p_X-p}={(1+z)r'-3r\over 3r-3\Omega_{\rm
m0}(1+z)^3}.
\end{equation}

Next, let us consider the generalized ghost condensate model
proposed in Ref.~\cite{Tsujikawa:2005ju} (see also
Ref.~\cite{Zhang:2006qu}), in which the behavior of crossing the
cosmological-constant boundary can be realized, with the Lagrangian
density
\begin{equation}
p=-X+h(\phi)X^2,
\end{equation}
where $h(\phi)$ is a function in terms of $\phi$. Actually, the
function $h(\phi)$ can be explicitly expressed for the specific
cases. For example, in the dilatonic ghost case, we have $h(\phi)=c
e^{\lambda\phi}$ \cite{ghost2}. From Eqs.~(\ref{p}) and (\ref{px})
we obtain
\begin{equation}
\phi'^2={12r-3(1+z)r'-3\Omega_{\rm m0}(1+z)^3\over r(1+z)^2},
\label{phip}
\end{equation}
\begin{equation}
h(\phi)={6(2(1+z)r'-6r+r(1+z)^2\phi'^2)\over
r^2(1+z)^4\phi'^4}\rho_{\rm c0}^{-1}, \label{hz}
\end{equation}
\begin{equation}
X=\frac{1}{2}\dot{\phi}^2=\frac{1}{6}r\phi'^2(1+z)^2\rho_{\rm c0},
\end{equation}
where $\rho_{\rm c0}=3H_0^2$ represents the present critical density
of the universe. The crossing of the cosmological-constant boundary
corresponds to $hX=1/2$. The system can enter the phantom region
($hX <1/2$) without discontinuous behavior of $h$ and $X$.

The evolution of the field $\phi$ can be derived by integrating
$\phi'$ according to Eq.(\ref{phip}). Note that the field $\phi$ is
determined up to an additive constant $\phi_0$, but it is convenient
to take $\phi$ to be zero at the present epoch ($z=0$). The function
$h(\phi)$ can be reconstructed using Eq.~(\ref{hz}) when the
information of $r(z)$ is obtained from the observational data.

Generically, the Friedmann equation can be expressed as
\begin{equation}
r(z)=\Omega_{\rm m0}(1+z)^3+(1-\Omega_{\rm m0})f(z),
\end{equation}
where $f(z)$ is some function encoding the information about the
dynamical property of dark energy,
\begin{equation}
f(z)=\exp{[3\int_{0}^{z}{1+w(s)\over 1+s}ds]}.
\end{equation}

\subsection{Reconstruction}

In this subsection, we will reconstruct the function $h(\phi)$ for
the ghost condensate model using some ansatzs for the
equation-of-state of dark energy. We will first use the WMAP5
parametrization discussed in section \ref{sec:para}. This case is
important in this paper because the ansatz is new. Next, for
comparing the new ansatz with the previous ones, we will preform the
same reconstruction program for other scenarios. This includes the
CPL parametrization and the holographic dark energy scenarios. The
reconstruction will correspond to the fitting results from the
latest observational data. What's more, the reconstruction program
will be implemented up to the decoupling epoch at $z_*\simeq 1090$,
which is different from the previous works
\cite{Tsujikawa:2005ju,Zhang:2006qu} that focus only on the late
times.

\subsubsection{WMAP5 parametrization}

\begin{figure}[htbp]
\begin{center}
\includegraphics[scale=0.75]{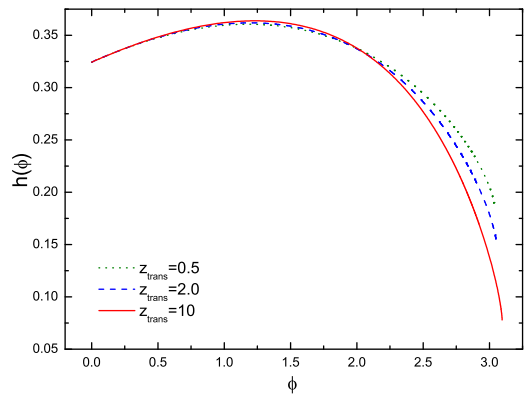}
\caption[]{\small Reconstruction of the generalized ghost condensate
model according to the WMAP5 parametrization with the best fit
results derived from WMAP5 combined with SN and BAO, $w_0=-1.06$,
$w'=0.36$ and $\Omega_{\rm m0}=0.273$. In this plot, we show the
cases of the function $h(\phi)$, in unit of $\rho_{\rm c0}^{-1}$.
The selected lines correspond to the transition redshift $z_{\rm
trans}=0.5$, $2.0$ and $10$, respectively.} \label{fig:hphi}
\end{center}
\end{figure}

\begin{figure}[htbp]
\begin{center}
\includegraphics[scale=0.75]{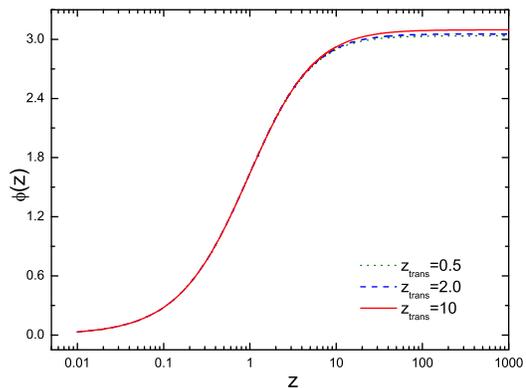}
\caption[]{\small Reconstruction of the generalized ghost condensate
model according to the WMAP5 parametrization with transition
redshift $z_{\rm trans}=0.5$, $2.0$ and $10$. In this plot, we show
the evolutions of the scalar field $\phi(z)$, in unit of the Planck
mass $M_{P}$, corresponding to the best fit results of the joint
analysis of WMAP5 $+$ SN $+$ BAO.}\label{fig:phiz}
\end{center}
\end{figure}

\begin{figure}[htbp]
\begin{center}
\includegraphics[scale=0.75]{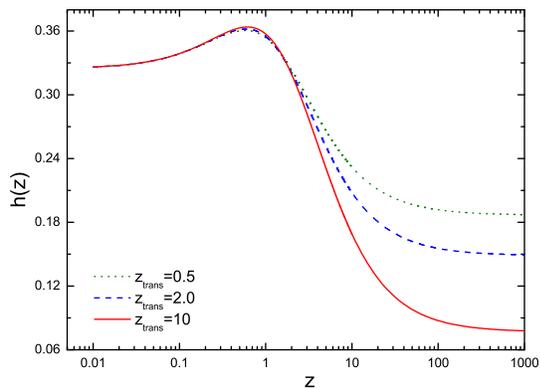}
\caption[]{\small Reconstruction of the generalized ghost condensate
model according to the WMAP5 parametrization with transition
redshift $z_{\rm trans}=0.5$, $2.0$ and $10$. In this plot, we show
the evolution of the function $h(z)$. Here $h$ is in unit of
$\rho_{\rm c0}^{-1}$.}\label{fig:hz}
\end{center}
\end{figure}

\begin{figure}[htbp]
\begin{center}
\includegraphics[scale=0.75]{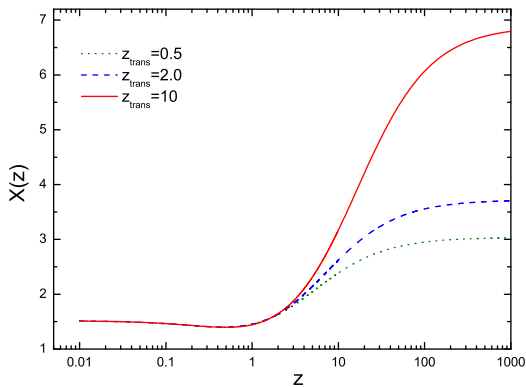}
\caption[]{\small Reconstruction of the generalized ghost condensate
model according to the WMAP5 parametrization with transition
redshift $z_{\rm trans}=0.5$, $2.0$ and $10$. In this plot, we show
the evolution of the kinematic energy density $X=\dot{\phi}^2/2$.
Here, $X$ is in unit of $\rho_{\rm c0}$.}\label{fig:xz}
\end{center}
\end{figure}

First, we use the new ansatz (\ref{eq:wz}) to implement the
reconstruction. The reconstruction for $h(\phi)$ is plotted in
Fig.~\ref{fig:hphi} with transition redshift $z_{\rm trans}=0.5$,
$2$ and $10$, by using the best-fit results, $w_0=-1.06$, $w'=0.36$
and $\Omega_{\rm m0}=0.273$, from the combined analysis of
WMAP5+SN+BAO. In addition, the evolutions of the scalar field
$\phi(z)$ as well as the functions $h(z)$ and $X(z)$ are also
determined by the reconstruction program, see Figs.~\ref{fig:phiz},
\ref{fig:hz} and \ref{fig:xz}.

From Fig.~\ref{fig:hphi}, we see that the reconstructed $h(\phi)$,
up to $z_*\simeq 1090$, is not a monotonous function. In the rough
range of $z$ between 0 and 1, the function $h(\phi)$ is increasing,
see also Fig.~\ref{fig:hz}. The shape of $h(\phi)$ in this range
indeed mimic an exponential function that is the case of the
dilatonic ghost condensate \cite{ghost2}. However, in the range of
$z$ greater than 1, $h(\phi)$ is a decreasing function.
Figure~\ref{fig:xz} shows the case of the kinematic energy density
$X(z)$. From this figure, we find that $z\simeq 1$ is indeed a pivot
point. In the range of $z$ larger than 1, the field $\phi$ moves
more and more slowly; in the range of $z$ less than 1, the field
$\phi$ moves faster and faster, albeit the change of $X$ in this
stage is slight. From Fig.~\ref{fig:phiz}, we can explicitly see the
change rate of the field $\phi$. We find that in the range of $z\sim
0.1-10$, the change rate of $\phi$, namely $d\phi/dz$, is large;
elsewhere, it is small.

One of the aims of this paper is to explore the dynamical evolution
behavior of the scalar field at early times (high redshifts), by
reconstructing the dynamics of the scalar field according to the
observations. Previous works only focus on the low redshift
evolution ($z<2$ or so) \cite{Tsujikawa:2005ju,Zhang:2006qu}. From
Figs.~\ref{fig:hz} and \ref{fig:xz}, we see that at low redshifts,
the cases with different $z_{\rm trans}$ behave in accordance, but
at high redshifts, the difference turns on. The bigger $z_{\rm
trans}$ is, the smaller $h$ and bigger $X$ are, at high redshifts.
For the scalar field evolution, we see from Fig.~\ref{fig:phiz} that
the difference in the shapes of $\phi(z)$ is not big. However, the
difference in shapes of $h(\phi)$ is rather evident for different
$z_{\rm trans}$. Therefore, our investigation of the reconstruction
explicitly exhibits the early-time dynamical evolution of the
generalized ghost condensate model. We show that, for the WMAP5
parametrization, different $z_{\rm trans}$ will bring little impact
at low redshifts but bring great impact at high redshifts, to the
dynamics of scalar field.

For a comparison, we shall also investigate other cases based on
different ansatzs or scenarios in what follows. In those cases, we
will only show the reconstructed $h(\phi)$ and $\phi(z)$, for
briefness.

\subsubsection{CPL parametrization}

We now consider the CPL ansatz for the equation-of-state of dark
energy, $w(a)=w_0+(1-a)w_a$. It should be pointed out that if one
extends it to an arbitrarily high redshift, it will result in an
undesirable situation in which the dark energy is as important as
the radiation density at the epoch of the Big Bang Nucleosynthesis
(BBN). Hence, in order to constrain such a scenario, one may use the
limit on the expansion rate from BBN.

The WMAP team also shows in Ref.~\cite{Komatsu:2008hk} the
constraint on $w_0$ and $w_a$ for the CPL model,
$w(a)=w_0+(1-a)w_a$, from the WMAP distance priors, the BAO and SN
data, and the BBN prior in the flat universe. The 95\% limit on
$w_0$ is $-0.29<1+w_0<0.21$ and the 68\% intervals are $w_0=-1.04\pm
0.13$ and $w_a=0.24\pm 0.55$. Besides, the effects of the systematic
errors are also studied. They find that $w_0=-1.00\pm 0.19$ and
$w_a=0.11\pm 0.70$ with the systematic errors included.

\begin{figure}[htbp]
\begin{center}
\includegraphics[scale=0.75]{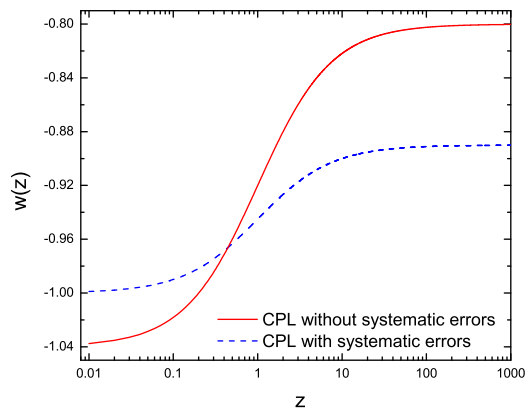}
\caption[]{\small The equation-of-state $w(z)$ in the CPL
parametrization. In this plot, we show the two best-fit cases from
WMAP5+SN+BAO+BBN, with and without SN systematic
errors.}\label{fig:cplwz}
\end{center}
\end{figure}

The dark energy equation-of-state of the two cases with and without
the SN systematic errors, for the CPL parametrization, at the
best-fits, is plotted in Fig.~\ref{fig:cplwz}. From this figure, one
can see that when considering the SN systematic errors, the fitting
results will be influenced significantly. One can find that the
equation-of-state even does not cross $-1$ in the CPL case with the
systematic errors, at the best-fit. Furthermore, comparing with the
WMAP5 parametrization (see Fig.~\ref{fig:wz}),  it is easy to see
that the early-time evolutionary behaviors for the equation-of-state
are very different.

\begin{figure}[htbp]
\begin{center}
\includegraphics[scale=0.75]{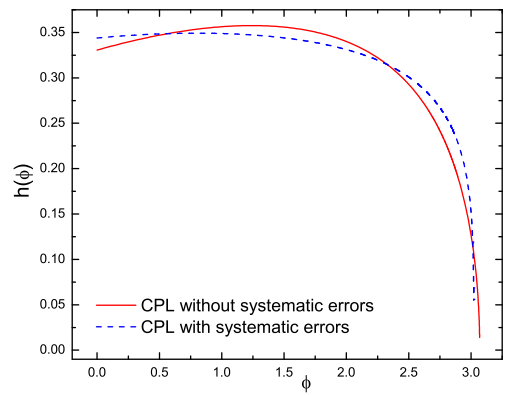}
\caption[]{\small Reconstruction of the generalized ghost condensate
model according to the CPL parametrization with the best fit results
derived from WMAP5 combined with SN, BAO, and BBN. The function
$h(\phi)$ is in unit of $\rho_{\rm c0}^{-1}$.} \label{fig:cplhphi}
\end{center}
\end{figure}

\begin{figure}[htbp]
\begin{center}
\includegraphics[scale=0.75]{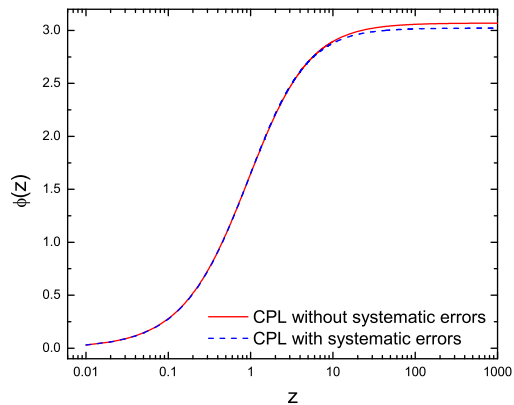}
\caption[]{\small Reconstruction of the generalized ghost condensate
model according to the CPL parametrization with the best fit results
derived from WMAP5 combined with SN, BAO, and BBN. The scalar field
$\phi(z)$ is in unit of $M_{P}$.}\label{fig:cplphiz}
\end{center}
\end{figure}

Performing the reconstruction program, we derive the function forms
of $h(\phi)$ and $\phi(z)$, shown in Figs.~\ref{fig:cplhphi} and
\ref{fig:cplphiz}, respectively. We find that the global trend of
the functions $h(\phi)$ and $\phi(z)$ of the CPL case is similar to
that of the WMAP5 case (see also Figs.~\ref{fig:hphi} and
\ref{fig:phiz}). For the function $h(\phi)$, comparing with the
WMAP5 case, the late-time behaviors are very similar but the
early-time behaviors are slightly different. Also, we find from
Fig.~\ref{fig:cplhphi} that the function $h(\phi)$ will be
monotonously decreasing if the equation-of-state does not cross $-1$
(see the dashed lines in Figs.~\ref{fig:cplwz} and
\ref{fig:cplhphi}). For the dynamical evolution of the field $\phi$,
comparing Fig.~\ref{fig:cplphiz} with Fig.~\ref{fig:phiz}, we find
that the difference is fairly little.

\subsubsection{Holographic dark energy scenarios}

Furthermore, we also consider the holographic dark energy scenarios.
The reason of considering the holographic dark energy is that we
should not only consider the simple parametrizations of dark energy,
but also consider some sophisticated dark energy models motivated by
quantum gravity.

The holographic dark energy density can be expressed as
\begin{equation}
\rho_{\rm de}=3c^2M_P^2L^{-2},
\end{equation}
where $c$ is a numerical parameter determined by observations, and
$L$ is the infrared (IR) cutoff of the theory. Here, we explicitly
write out the reduced Planck mass $M_P$. In the holographic dark
energy models, the key problem is how to choose an appropriate IR
cutoff for the theory. In the original holographic dark energy
scenario proposed by Li \cite{holo}, the IR cutoff is chosen as the
event horizon of the universe, $R_{\rm eh}=a\int_t^\infty dt/a$. In
a generalized version \cite{Gao:2007ep}, the IR cutoff is taken as
the average of the Ricci scalar curvature, $|{\cal R}|^{-1/2}$. This
new version is often called ``Ricci dark energy.'' It should be
mentioned that the two scenarios of holographic dark energy both
exhibit quintom feature \cite{holo,Gao:2007ep,RDERS}.

\begin{figure}[htbp]
\begin{center}
\includegraphics[scale=0.75]{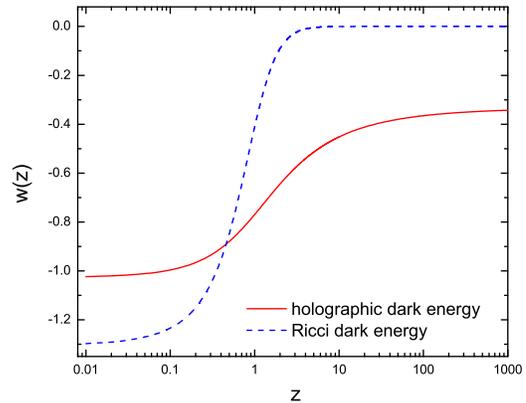}
\caption[]{\small The equation-of-state $w(z)$ in the holographic
dark energy scenarios. In this plot, we show the best-fit cases from
WMAP5+SN+BAO.}\label{fig:holowz}
\end{center}
\end{figure}

Recently, the holographic dark energy models were constrained by the
latest observational data, WMAP5+BAO+SN, see Ref.~\cite{Li:2009bn}.
For the holographic dark energy, we have the fitting results: For
$68.3\%$ confidence level, $\Omega_{\rm
m0}=0.277^{+0.022}_{-0.021}$, and $c=0.818^{+0.113}_{-0.097}$; for
$95.4\%$ confidence level, $\Omega_{\rm
m0}=0.277^{+0.037}_{-0.034}$, and $c=0.818^{+0.196}_{-0.154}$. For
the Ricci dark energy, we have the fitting results: For $68.3\%$
confidence level, $\Omega_{\rm m0}=0.324^{+0.024}_{-0.022}$, and
$c^2=0.371^{+0.023}_{-0.023}$; for $95.4\%$ confidence level,
$\Omega_{\rm m0}=0.324^{+0.040}_{-0.036}$, and
$c^2=0.371^{+0.037}_{-0.038}$. The dark-energy equation of state, at
the best fits, in these two scenarios is shown in
Fig.~\ref{fig:holowz}. One can see from this figure that although
both originated from the holographic principle of quantum gravity,
different IR cutoffs will bring so different cosmological
consequences. We shall make use of the best-fit results to
reconstruct the ghost condensate model in the following.

\begin{figure}[htbp]
\begin{center}
\includegraphics[scale=0.75]{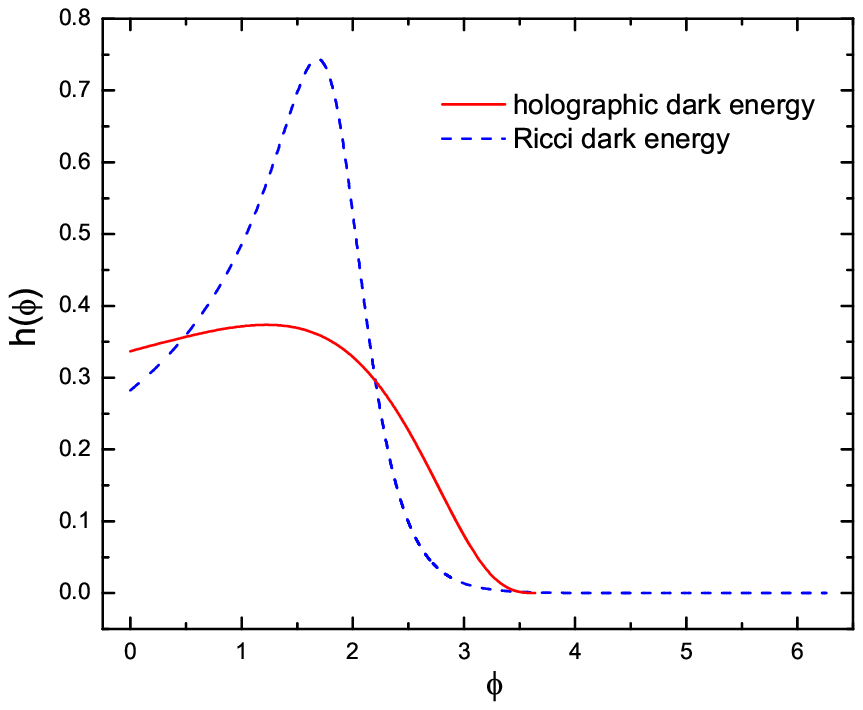}
\caption[]{\small Reconstruction of the generalized ghost condensate
model according to the holographic dark energy scenarios with the
best-fit results derived from WMAP5 combined with SN and BAO. The
function $h(\phi)$ is in unit of $\rho_{\rm c0}^{-1}$.}
\label{fig:holohphi}
\end{center}
\end{figure}

\begin{figure}[htbp]
\begin{center}
\includegraphics[scale=0.75]{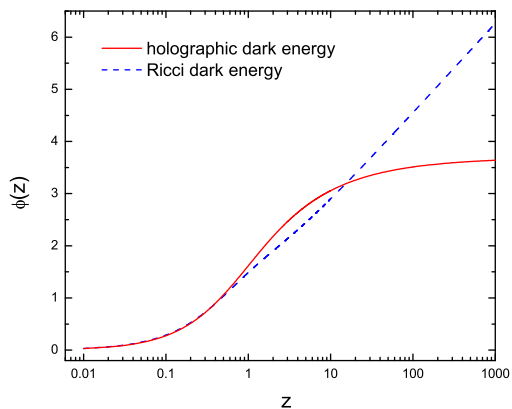}
\caption[]{\small Reconstruction of the generalized ghost condensate
model according to the holographic dark energy scenarios with the
best-fit results derived from WMAP5 combined with SN and BAO. The
scalar field $\phi(z)$ is in unit of $M_{P}$.}\label{fig:holophiz}
\end{center}
\end{figure}

The reconstructed function forms of $h(\phi)$ and $\phi(z)$ are
shown in Figs.~\ref{fig:holohphi} and \ref{fig:holophiz}. From these
two figures, we see that the big difference in $w(z)$ is converted
to the big differences in $h(\phi)$ and $\phi(z)$. The
reconstructions of $h(\phi)$ and $\phi(z)$ indicate that the
holographic dark energy is compatible with the previous dark energy
parametrizations, but the Ricci dark energy is not. In
Fig.~\ref{fig:holohphi}, we find that there exist a sharp peak of
$h$ around $\phi\sim 1.5$ and a long tail of $h$ in the range of
$\phi>3.5$, for the Ricci dark energy. From Fig.~\ref{fig:holophiz},
we see that the dynamics of the field $\phi$ in the Ricci scenario
is also larruping. Although in the range of $z<1$, the evolutions of
$\phi$ nearly go to degenerate, the big different occurs in the
range of $z>1$, especially in the stage of $z>10$.

In Ref.~\cite{Li:2009bn}, the authors use the Bayesian evidence (BE)
as a model selection criterion to make a comparison between the
holographic dark energy models. It is found that for holographic
dark energy and Ricci dark energy, $\Delta \ln \mathrm{BE}= -0.86$
and $-8.14$, respectively. So, evidently, the holographic dark
energy scenrio is more favored by the observational data, whereas
the Ricci dark energy scenario looks like disfavored by the
observational data. Our reconstruction investigation also supports
this conclusion from another angle of view.

\section{Concluding remarks}\label{sec:concl}

The recent fits to current observational data, such as SN, CMB and
LSS, find that even though the behavior of dark energy is consistent
to great extent with a cosmological constant, an evolving dark
energy with the equation of state $w$ larger than $-1$ in the recent
past but less than $-1$ today is also with some possibility.
Although the scalar-field models of dark energy, such as
quintessence and phantom, can provide us with dynamical mechanism
for dark energy, the behavior of cosmological-constant crossing
brings forward a great challenge to the model-building for dynamical
dark energy, because neither quintessence nor phantom can fulfill
this behavior. A two-field quintom model, therefore, was suggested
to realize this behavior by means of the incorporation of the
features of quintessence and phantom. Besides, the generalized ghost
condensate model provides us with a successful
single-real-scalar-field model for realizing the quintom-like
behavior. For probing the dynamical nature of dark energy, one
should parameterize dark energy first and then constrain the
parameters using the observational data.

In this paper, we have investigated the dynamical behavior of the
general ghost condensate scalar-field model by taking a new form of
parametrization of the equation of state proposed in
Ref.~\cite{Komatsu:2008hk} and the best-fit values from the
observational data. The results of reconstruction show the dynamical
behavior of the generalized ghost condensate from this
parametrization. In particular, this reconstruction investigation
explores the early-time evolutionary behavior of the scalar field
model. As a comparison, we also discussed other specific cases
including the CPL parametrization and the holographic dark energy
scenarios.

The increase of the quantity and quality of observational data in
the future will undoubtedly provide a true {\it model-independent}
manner for exploring the properties of dark energy. We hope that the
future high-precision observations (e.g. SNAP) may be capable of
providing us with deep insight into the nature of dark energy
driving the acceleration of the universe.

\section*{ACKNOWLEDGMENTS}

We thank Xin Zhang for helpful discussions. This work was supported
by the Natural Science Foundation of China under Grants
Nos.~10705041 and 10975032.



\begin{thebibliography}{99}


\bibitem{SN}
  A.~G.~Riess {\it et al.}  [Supernova Search Team Collaboration],
  Astron.\ J.\  {\bf 116}, 1009 (1998)
  [astro-ph/9805201];
  S.~Perlmutter {\it et al.}  [Supernova Cosmology Project Collaboration],
  Astrophys.\ J.\  {\bf 517}, 565 (1999)
  [astro-ph/9812133].

\bibitem{LSS}
  M.~Tegmark {\it et al.}  [SDSS Collaboration],
  Phys.\ Rev.\ D {\bf 69}, 103501 (2004)
  [astro-ph/0310723];
  K.~Abazajian {\it et al.}  [SDSS Collaboration],
  Astron.\ J.\  {\bf 128}, 502 (2004)
  [astro-ph/0403325];
  K.~Abazajian {\it et al.}  [SDSS Collaboration],
  Astron.\ J.\  {\bf 129}, 1755 (2005)
  [astro-ph/0410239].



\bibitem{CMB}
  D.~N.~Spergel {\it et al.}  [WMAP Collaboration],
  Astrophys.\ J.\ Suppl.\  {\bf 148}, 175 (2003)
  [astro-ph/0302209].


\bibitem{Einstein:1917} A. Einstein, Sitzungsber. K. Preuss.
Akad. Wiss. 142 (1917).

\bibitem{coincidence} P.~J.~Steinhardt, in {\it Critical Problems in
Physics}, edited by V.~L.~Fitch and D.~R.~Marlow (Princeton
University Press, Princeton, NJ, 1997).

\bibitem{cc}S.~Weinberg,
Rev.\ Mod.\ Phys.\  {\bf 61} 1 (1989).




\bibitem{quintessence}
  P.~J.~E.~Peebles and B.~Ratra,
  Astrophys.\ J.\  {\bf 325} L17 (1988);
  B.~Ratra and P.~J.~E.~Peebles,
  Phys.\ Rev.\ D {\bf 37} 3406 (1988);
  C.~Wetterich,
  Nucl.\ Phys.\ B {\bf 302} 668 (1988);
  J.~A.~Frieman, C.~T.~Hill, A.~Stebbins and I.~Waga,
  Phys.\ Rev.\ Lett.\  {\bf 75}, 2077 (1995)
  [astro-ph/9505060];
  M.~S.~Turner and M.~J.~White,
  Phys.\ Rev.\ D {\bf 56}, 4439 (1997)
  [astro-ph/9701138];
  A.~R.~Liddle and R.~J.~Scherrer,
  Phys.\ Rev.\ D {\bf 59}, 023509 (1999)
  [astro-ph/9809272];
  I.~Zlatev, L.~M.~Wang and P.~J.~Steinhardt,
  Phys.\ Rev.\ Lett.\  {\bf 82}, 896 (1999)
  [astro-ph/9807002];
  P.~J.~Steinhardt, L.~M.~Wang and I.~Zlatev,
  Phys.\ Rev.\ D {\bf 59}, 123504 (1999)
  [astro-ph/9812313];
  X.~Zhang,
  Mod.\ Phys.\ Lett.\ A {\bf 20}, 2575 (2005)
  [astro-ph/0503072];
  X.~Zhang,
  Phys.\ Lett.\ B {\bf 611}, 1 (2005)
  [astro-ph/0503075].


\bibitem{kessence}
  C.~Armendariz-Picon, V.~F.~Mukhanov and P.~J.~Steinhardt,
  Phys.\ Rev.\ Lett.\  {\bf 85}, 4438 (2000)
  [astro-ph/0004134];
  C.~Armendariz-Picon, V.~F.~Mukhanov and P.~J.~Steinhardt,
  Phys.\ Rev.\ D {\bf 63}, 103510 (2001)
  [astro-ph/0006373].



\bibitem{phantom}
  R.~R.~Caldwell,
  Phys.\ Lett.\ B {\bf 545}, 23 (2002)
  [astro-ph/9908168].


\bibitem{tachyon}
  A.~Sen,
  JHEP {\bf 0207}, 065 (2002)
  [hep-th/0203265];
  T.~Padmanabhan,
  Phys.\ Rev.\ D {\bf 66}, 021301 (2002)
  [hep-th/0204150].

\bibitem{ghost1}
  N.~Arkani-Hamed, H.~C.~Cheng, M.~A.~Luty and S.~Mukohyama,
  JHEP {\bf 0405}, 074 (2004)
  [hep-th/0312099];
  S.~Mukohyama,
  JCAP {\bf 0610}, 011 (2006)
  [hep-th/0607181].



\bibitem{ghost2}
  F.~Piazza and S.~Tsujikawa,
  JCAP {\bf 0407}, 004 (2004)
  [hep-th/0405054].

\bibitem{Vikman:2004dc}
  A.~Vikman,
  Phys.\ Rev.\ D {\bf 71}, 023515 (2005)
  [astro-ph/0407107].


\bibitem{quintom}
  B.~Feng, X.~L.~Wang and X.~M.~Zhang,
  Phys.\ Lett.\ B {\bf 607}, 35 (2005)
  [astro-ph/0404224].


\bibitem{twofield}
  Z.~K.~Guo, Y.~S.~Piao, X.~M.~Zhang and Y.~Z.~Zhang,
  Phys.\ Lett.\ B {\bf 608}, 177 (2005)
  [astro-ph/0410654];
  X.~Zhang,
  Commun.\ Theor.\ Phys.\  {\bf 44}, 762 (2005);
  X.~F.~Zhang, H.~Li, Y.~S.~Piao and X.~M.~Zhang,
  Mod.\ Phys.\ Lett.\ A {\bf 21}, 231 (2006)
  [astro-ph/0501652].


\bibitem{quintom2}
  Y.~F.~Cai, H.~Li, Y.~S.~Piao and X.~M.~Zhang,
  Phys.\ Lett.\  B {\bf 646}, 141 (2007)
  [arXiv:gr-qc/0609039];
  Y.~F.~Cai, T.~Qiu, Y.~S.~Piao, M.~Li and X.~M.~Zhang,
  JHEP {\bf 0710}, 071 (2007)
  [arXiv:0704.1090 [gr-qc]];
  H.~H.~Xiong, Y.~F.~Cai, T.~Qiu, Y.~S.~Piao and X.~M.~Zhang,
  Phys.\ Lett.\  B {\bf 666}, 212 (2008)
  [arXiv:0805.0413 [astro-ph]];
  R.~Lazkoz, G.~Leon and I.~Quiros,
  Phys.\ Lett.\  B {\bf 649}, 103 (2007)
  [arXiv:astro-ph/0701353];
  H.~Wei and R.~G.~Cai,
  Phys.\ Lett.\  B {\bf 634}, 9 (2006)
  [arXiv:astro-ph/0512018];
  S.~G.~Shi, Y.~S.~Piao and C.~F.~Qiao,
  JCAP {\bf 0904}, 027 (2009)
  [arXiv:0812.4022 [astro-ph]];
  M.~R.~Setare and E.~N.~Saridakis,
  arXiv:0807.3807 [hep-th];
  L.~P.~Chimento, M.~Forte, R.~Lazkoz and M.~G.~Richarte,
  Phys.\ Rev.\  D {\bf 79}, 043502 (2009)
  [arXiv:0811.3643 [astro-ph]].






\bibitem{Anisimov:2005ne}
  A.~Anisimov, E.~Babichev and A.~Vikman,
  JCAP {\bf 0506}, 006 (2005)
  [astro-ph/0504560].

\bibitem{Li:2005fm}
  M.~Z.~Li, B.~Feng and X.~M.~Zhang,
  JCAP {\bf 0512}, 002 (2005)
  [hep-ph/0503268].

\bibitem{Tsujikawa:2005ju}
  S.~Tsujikawa,
  Phys.\ Rev.\ D {\bf 72}, 083512 (2005)
  [astro-ph/0508542].

\bibitem{Zhang:2006qu}
  X.~Zhang,
  Phys.\ Rev.\ D {\bf 74}, 103505 (2006)
  [astro-ph/0609699];
  X.~Zhang,
  Phys. Rev. D {\bf 79}, 103509 (2009)
  [arXiv:0901.2262 [astro-ph.CO]];
  J.~Zhang, X.~Zhang and H.~Liu,
  Mod.\ Phys.\ Lett.\  A {\bf 23}, 139 (2008)
  [arXiv:astro-ph/0612642];
  C.~J.~Feng,
  Phys.\ Lett.\  B {\bf 672}, 94 (2009)
  [arXiv:0810.2594 [hep-th]].






\bibitem{Huang:2005gu}
  C.~G.~Huang and H.~Y.~Guo,
  astro-ph/0508171.

\bibitem{holo}
  M.~Li,
  Phys.\ Lett.\ B {\bf 603}, 1 (2004)
  [hep-th/0403127];
  Q.~G.~Huang and M.~Li,
  JCAP {\bf 0408}, 013 (2004)
  [astro-ph/0404229];
  Q.~G.~Huang and Y.~G.~Gong,
  JCAP {\bf 0408}, 006 (2004)
  [astro-ph/0403590];
  Q.~G.~Huang and M.~Li,
  JCAP {\bf 0503}, 001 (2005)
  [hep-th/0410095];
  X.~Zhang,
  Int.\ J.\ Mod.\ Phys.\ D {\bf 14}, 1597 (2005)
  [astro-ph/0504586];
  X.~Zhang and F.~Q.~Wu,
  Phys.\ Rev.\ D {\bf 72}, 043524 (2005)
  [astro-ph/0506310];
  Z.~Chang, F.~Q.~Wu and X.~Zhang,
  Phys.\ Lett.\ B {\bf 633}, 14 (2006)
  [astro-ph/0509531];
  X.~Zhang and F.~Q.~Wu,
  Phys.\ Rev.\  D {\bf 76}, 023502 (2007)
  [arXiv:astro-ph/0701405];
  J.~Zhang, X.~Zhang and H.~Liu,
  Eur.\ Phys.\ J.\  C {\bf 52}, 693 (2007)
  [arXiv:0708.3121 [hep-th]];
  M.~R.~Setare, J.~Zhang and X.~Zhang,
  JCAP {\bf 0703}, 007 (2007)
  [arXiv:gr-qc/0611084];
  J.~Zhang, X.~Zhang and H.~Liu,
  Phys.\ Lett.\  B {\bf 659}, 26 (2008)
  [arXiv:0705.4145 [astro-ph]];
  M.~Li, X.~D.~Li, C.~S.~Lin and Y.~Wang,
  Commun.\ Theor.\ Phys.\  {\bf 51}, 181 (2009)
  [arXiv:0811.3332 [hep-th]];
  X.~Zhang,
  arXiv:0909.4940 [gr-qc].


\bibitem{Cai:2005ie}
  R.~G.~Cai, H.~S.~Zhang and A.~Wang,
  Commun.\ Theor.\ Phys.\  {\bf 44}, 948 (2005)
  [hep-th/0505186].

\bibitem{simplescalar}
A. A. Starobinsky, JETP Lett. {\bf68}, 757 (1998)
[astro-ph/9810431]; D. Huterer and M. S. Turner, Phys. Rev. D
{\bf60}, 081301 (1999) [astro-ph/9808133]; T. Nakamura and T. Chiba,
Mon. Not. R. Astron. Soc. {\bf306}, 696 (1999) [astro-ph/9810447];
Z.~K.~Guo, N.~Ohta and Y.~Z.~Zhang,
  Phys.\ Rev.\ D {\bf 72}, 023504 (2005)
  [astro-ph/0505253];
  X.~Zhang,
  Phys.\ Lett.\  B {\bf 648}, 1 (2007)
  [arXiv:astro-ph/0604484];
  Z.~K.~Guo, N.~Ohta and Y.~Z.~Zhang,
  astro-ph/0603109;
  Y.~Z.~Ma and X.~Zhang,
  Phys.\ Lett.\  B {\bf 661}, 239 (2008)
  [arXiv:0709.1517 [astro-ph]].




\bibitem{Saini:1999ba}
  T.~D.~Saini, S.~Raychaudhury, V.~Sahni and A.~A.~Starobinsky,
  Phys.\ Rev.\ Lett.\  {\bf 85}, 1162 (2000)
  [astro-ph/9910231].

\bibitem{scalartensor}
B. Boisseau, G. Esposito-Farese, D. Polarski and A. A. Starobinsky,
Phys. Rev. Lett. {\bf85}, 2236 (2000) [gr-qc/0001066]; G.
Esposito-Farese and D. Polarski, Phys. Rev. D {\bf63}, 063504 (2001)
[gr-qc/0009034]; L. Perivolaropoulos, JCAP {\bf0510}, 001 (2005)
[astro-ph/0504582].

\bibitem{frgrav}
S. Capozziello, V.F. Cardone and A. Troisi, Phys. Rev. D {\bf71},
043503 (2005) [astro-ph/0501426];
  X.~Wu and Z.~H.~Zhu,
  Phys.\ Lett.\  B {\bf 660}, 293 (2008)
  [arXiv:0712.3603 [astro-ph]];
  C.~J.~Feng,
  arXiv:0812.2067 [hep-th].





\bibitem{Li:2006bx}
  H.~Li, Z.~K.~Guo and Y.~Z.~Zhang,
  astro-ph/0601007.

\bibitem{Gao:2007ep}
  C.~Gao, F. Q. Wu, X.~Chen and Y.~G.~Shen,
  Phys. Rev. D {\bf 79}, 043511 (2009)
  [arXiv:0712.1394 [astro-ph]].



\bibitem{holotach}
  J.~Zhang, X.~Zhang and H.~Liu,
  Phys.\ Lett.\  B {\bf 651}, 84 (2007)
  [arXiv:0706.1185 [astro-ph]];
  J.~Cui, L.~Zhang, J.~Zhang and X.~Zhang,
  arXiv:0902.0716 [astro-ph.CO].




\bibitem{Komatsu:2008hk}
  E.~Komatsu {\it et al.}  [WMAP Collaboration],
  Astrophys.\ J.\ Suppl.\  {\bf 180}, 330 (2009)
  [arXiv:0803.0547 [astro-ph]].

\bibitem{DiPietro:2002cz}
E.~Di Pietro and J.~F.~Claeskens,
 Mon.\ Not.\ Roy.\ Astron.\ Soc.\  {\bf 341}, 1299 (2003)
 [astro-ph/0207332].

\bibitem{Riess:2004nr}
  A.~G.~Riess {\it et al.}  [Supernova Search Team Collaboration],
 Astrophys.\ J.\  {\bf 607}, 665 (2004)
 [astro-ph/0402512].

\bibitem{Chevallier:2000qy}
 M.~Chevallier and D.~Polarski,
 Int.\ J.\ Mod.\ Phys.\ D {\bf 10}, 213 (2001)
 [gr-qc/0009008].



\bibitem{Linder:2002et}
 E.~V.~Linder,
 Phys.\ Rev.\ Lett.\  {\bf 90}, 091301 (2003)
 [astro-ph/0208512].



\bibitem{Wang:2007mza}
  Y.~Wang and P.~Mukherjee,
  Phys.\ Rev.\  D {\bf 76}, 103533 (2007)
  [arXiv:astro-ph/0703780].

\bibitem{RDERS}
  C.~J.~Feng and X.~Zhang,
  Phys.\ Lett.\  B {\bf 680}, 399 (2009)
  [arXiv:0904.0045 [gr-qc]].



\bibitem{Li:2009bn}
  M.~Li, X.~D.~Li, S.~Wang and X.~Zhang,
  JCAP {\bf 0906}, 036 (2009)
  [arXiv:0904.0928 [astro-ph.CO]].











\end{thebibliography}
\end{document}